# IMPLEMENTATION OF LINEAR DETECTION TECHNIQUES TO OVERCOME CHANNEL EFFECTS IN MIMO

Gopika k [1] and M Mathurakani[2]

[1]Student in M.Tech, ECE Department, Toc H Institute of Science and Technology, Ernakulam, India
[2]Professor, ECE Department, Toc H Institute of Science and Technology
Formerly Scientist, DRDO, NPOL, Kochi,Ernakulam, India

**ABSTRACT**

*Spatial diversity technique enables improvement in quality and reliability of wireless link. Antenna diversity along with understanding effects of channel on transmitted signal and methods to overcome the channel impairment plays an important role in wireless communication where sharing of channel occurs between users. In this paper single input single output system (SISO) is compared with multiple input multiple output system (MIMO) in terms of bit error rate performance. Bit error rate performance is also evaluated for MIMO with least squares (LS) and Minimum mean square error (MMSE) linear detection. Further analysis and simulation is done to understand the effect of channel imperfections on BER.*

## 1. INTRODUCTION

Today's and future wireless communication devices are expected to support various multimedia services and demands a huge bandwidth, but it is a scarce resource and hence needed to be managed and utilized carefully. To effectively utilize this limited spectrum different users have to share the channel. The communication takes place in same space and uses same spectrum, interference will occur as a result and is a major limitation in wireless communication. So interference needs to be managed effectively, otherwise it will affect the system performance and limits the capacity that the system can achieve.

Evolution of antenna terminology started from transmitter and receiver equipped with single antennas for transmitting and receiving [1]. Achievable capacity was limited and the system known as SISO worked best for only line of sight distances. As a communication scenario is concerned not only line of sight but non line of sight communication also takes place. To achieve better capacity, spatial diversity techniques were implemented thus evolved multiple antenna terminals at transmitter called as multiple input single output (MISO) system enabling transmit diversity and multiple antenna terminals at receiver called single input multiple output (SIMO) system enabling receive diversity. These systems even though worked better when comparing with SISO, capacity achieved was poor. It necessitates the importance of achieving diversity at both transmitter and receiver leading to the MIMO systems, where in this system the capacity achieved depends on the minimum number of transmit or receive antennas [2]. Significant advantages of MIMO systems are increase in both system capacity and spectral efficiency. The capacity of a wireless link increases linearly with the minimum number of transmitter or receiver antennas. The data rate can be increased by spatial multiplexing without consuming more frequency resources and without increasing total transmitter power. Reduction of effects of fading due to increased diversity is particularly beneficial when different channels fade independently. MIMO system enabled joint processing or combining of signals and system integrity. A single MIMO system is called single user MIMO, because the single user terminal with which both transmit and receive diversity is enabled. In communication space not a single user but large numbers of users are present leading to the evolution of multi user MIMO. Therefore future works in this project will be focusing towards multiuser MIMO and interference alignment techniques.

The quality of a wireless link can be described by three parameters, namely the transmission rate, transmission range and transmission reliability. Transmission rate can be increased by reducing transmission range and reliability. By reducing transmission rate and reliability transmission range can be increased, while transmission reliability can be increased by reducing transmission rate and range. By combining two important technologies MIMO technology and orthogonal frequency division multiplexing (OFDM) above parameters can be simultaneously improved.

## 2. ORTHOGONAL FREQUENCY DIVISION MULTIPLEXING

In single carrier modulation the transmitted pulse width has to be reduced as it is occupying more bandwidth. If width of transmitted pulse is reduced channel which was narrow band channel initially starts behaving like a wideband channel resulting in severe inter symbol interference, so single carrier modulation is not effective. When a signal propagates through a mobile radio channel the transmitted signal will undergo a variation in its characteristics like amplitude, phase etc. referred to as fading of a signal. Multipath phenomenon will generate two effects mainly, frequency selective fading and intersymbol interference. Frequency selective fading occurs when the signal is transmitted through a constant gain channel with linear phase response over a bandwidth that is smaller than the bandwidth of transmitted signal. When a signal undergoes frequency selective fading at the receiver the received signal will get distorted and multiple versions of transmitted signal will be received which is attenuated and delayed in time i.e. for some frequencies in the bandwidth the channel does not allow any information to go through and thus deep fades occurs to particular frequencies. It does not occur uniformly across the band but occurs at selected frequencies. OFDM is implemented to overcome these impairments that the signal is suffering from when transmitted through a shared channel [3].

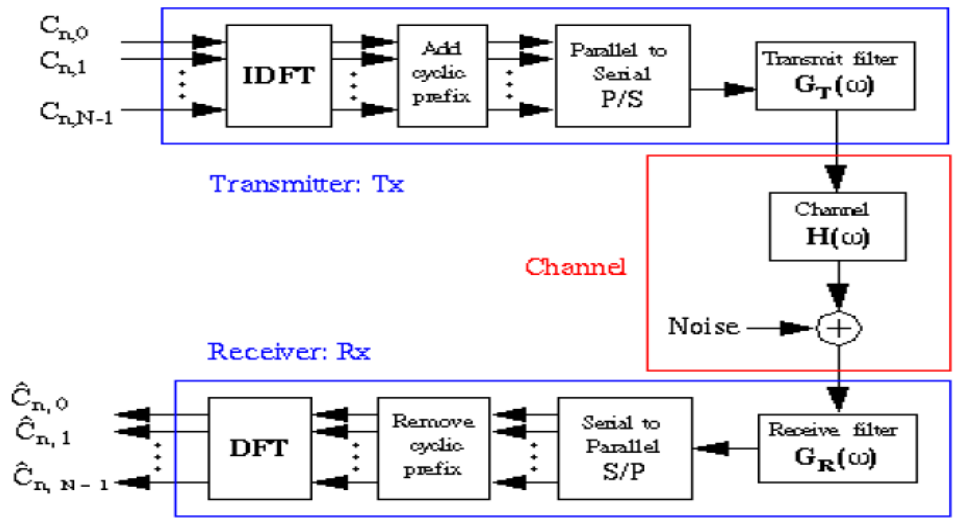

Figure 1. OFDM system

OFDM is a discrete implementation of multicarrier modulation; it divides the entire bit stream to be transmitted in to sub streams and sends them over different subchannels. The subchannels are orthogonal and the number of subchannels is designed such that each subchannel has a bandwidth less than the coherence bandwidth of channel.

Figure 1.shows the block diagram of OFDM system. The incoming bits are mapped to symbols according to some modulation scheme like quadrature phase shift keying (QPSK). The serial data is converted in to parallel blocks, then each block of symbols is forwarded to inverse fast fourier transform (IFFT) block and OFDM modulated. Then the OFDM signal will be appended with a cyclic prefix by copying last portion of OFDM signal. The cyclic prefix length is chosen such that its length should be larger than the maximum path delay of the channel, to eliminate intersymbol interference (ISI). Then the serially converted OFDM signal is transmitted. At the channel this signal will undergo some transformation like convolution with the channel; the linear convolution at the channel will be converted in to circular convolution due to the presence of cyclic prefix. At the receiver a reverse process that took place in transmitter occurs, the signal is converted in to parallel signals and cyclic prefix is removed, fast fourier transform (FFT) of signal is taken and channel effect can be removed by simply dividing this signal FFT with channel FFT, an advantage of cyclic prefix and hence circular convolution is that circular convolution in time domain will be converted in to multiplication in frequency domain, by doing this frequency selective fading channel is converted in to flat fading channel in subchannel perspective. OFDM is easy to implement in digital domain, its bandwidth efficient, robust to fading and is flexible in resource allocation.

## 3. MIMO SYSTEM MODEL

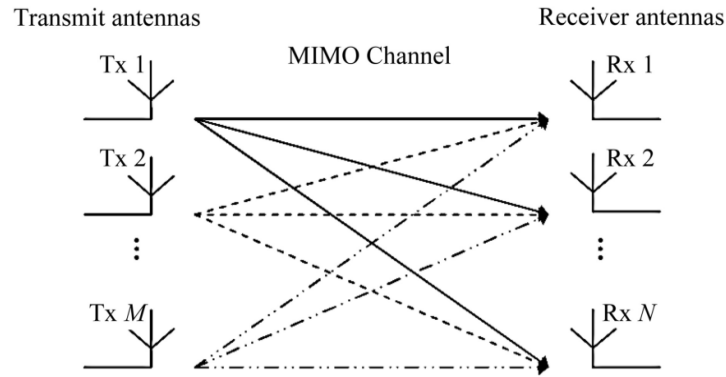

Figure 2. MIMO system

### 3.1 MIMO

Multiple input multiple output is a method implemented to improve the system capacity and capacity of a radio link using multiple antenna terminals at both transmitter and receiver side by utilizing multipath propagation. Through the multiple transmitting antennas either multiple copies of transmitted data streams to enhance the diversity gain or multiple data streams to improve the spatial multiplexing gain can be transmitted. When the multiple copies of data streams are transmitted from different terminals each stream will choose different paths to its receivers, during its propagation some of the data streams undergoes deep fading but other streams through different path survives and can be received at the receiver antenna with diversity enabled and datas can be recovered. Figure 2.illustrates a MIMO system with M transmitting and N receiving antennas and a MIMO channel in between them, the size of channel matrix is NxM

### 3.2 Channel State Information

As the data streams S passes through the channel H, signal undergo some variation and noise N will get added in to it. So the received stream Y will be,

Y=HxS+N.                                                                                                                        (1)

Receiver can suppress the effects of noise by increasing signal to noise ratio. But to deal with H receiver needs to have the knowledge about channel. Receiver has to be simple in terms of cost and size, but if the channel is predicted at the receiver and thus detecting data will increase the complexity of receiver. So the transmitter will do the hard work of predicting the channel and sending it along with the data streams through precoding. If the channel assumed is imperfect, error rate will increase and received signal will be distorted. Channel state information can be acquired through assuming reciprocity of alignment, it's the signalling dimensions along which a receiving node sees the least interference from other users are the signalling dimensions along which this node will cause least interference to other nodes in reciprocal network where all transmitters and receivers switch roles.

## 4. SPACE TIME CODING AND SPACE FREQUENCY CODING

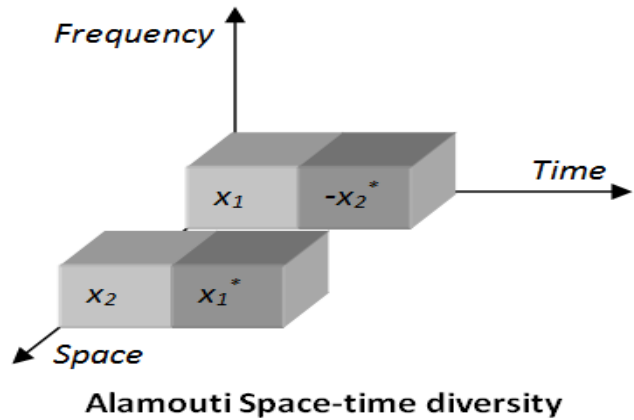

Figure 3. STBC illustration

Space time block codes (STBC) are spatial temporal codes that use diversity enabled in transmitter and receiver, multiple copies of data are transmitted in the hope that at least some of them may survive the physical transmission path. Transmit diversity proposed by alamouti was the first space time block codes [4]. Figure 3 shows the space time coding implementation using two transmit and two receive antennas. STBC in particular, the data stream to be transmitted is encoded in blocks, which are distributed among spaced antennas and across time. While it is necessary to have multiple transmit antennas, it is not necessary to have multiple receive antennas, although to do so improves performance. This process of receiving diverse copies of the data is known as diversity reception. An STBC is usually represented by a matrix. Each row represents a time slot and each column represents one antenna's transmissions over time. Two symbols X1 and X2 are transmitted from transmitter 1 and 2 at time slot 2n then at timeslot 2n+1,-X2*,X1* are transmitted from two transmitters to enable diversity. Simple linear operations are performed at the receiver such that X1 should be received at receiver1 and X2 at receiver2.

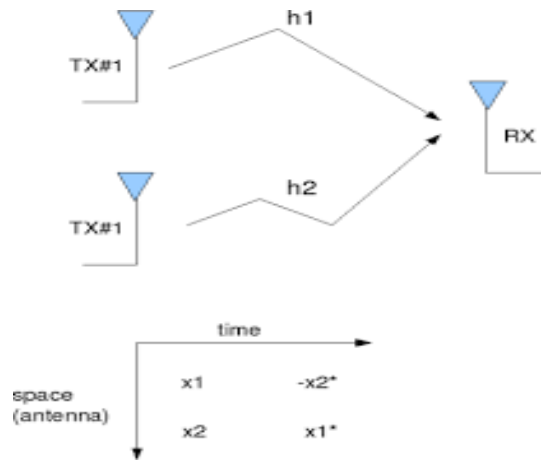

Figure 4. Alamouti encoding

Figure 4.shows the signals received at two receivers at two different slots 2n and 2n+1 can be represented by set of equations

Y1(2n)=H11xX1+H12xX2.                                                                                                  (2)
Y2(2n)=H21xX1+H22xX2.                                                                                                  (3)
Y1(2n+1)=-H11xconj(X2)+H12xconj (X1).                                                                         (4)
Y2(2n+1)=-H21xX2+H22xconj (X1).                                                                                    (5)

Hij is the channel response between j$^{th}$ transmitter and i$^{th}$ receiver. Y1 and Y2 are the symbols received at receiver 1 and 2 respectively. Here two transmit and receive antennas are used for simple implementation; higher order diversity can be enabled by increasing the number of antennas. The symbols can be retrieved at receiver

X1=conj(H11)xY1(2n)+H12xconj(Y1(2n+1)) +conj(H21)xY2(2n)+H22xconj(2n+1).       (6)
X2=conj(H12)xY1(2n)+H11xconj(Y1(2n+1)) +conj(H22)xY2(2n)-H22xconj(2n+1).        (7)

Space frequency coded alamouti transmission scheme is over different frequency rather than over different time slots as in space time coding, where a symbol goes through four different paths at two different frequencies thus achieving frequency and space diversity [5].

## 5. LEAST SQUARES AND MMSE LINEAR DETECTION

The transmitted signal after passing through the channel will undergo Rayleigh fading and additive white Gaussian noise will get added to the signal. So to remove the channel effect LS or MMSE based linear detection can be used.

In LS the linear detection is done as
S=(HxH$^H$ )$^{-1}$xH$^H$xY.                                                                                                                  (8)

In MMSE the linear detection is done as

188

International Journal on Cybernetics & Informatics (IJCI) Vol. 5, No. 2, April 2016

$S = (HxH^H + \sigma^2)^{-1} xH^H xY.$  (9)

H is the channel matrix; Y is the received symbol vector; $\sigma^2$ is the noise variance. MMSE considers effect of noise in linear detection procedure and is assumed to perform better than LS based linear detection.

## 6. SIMULATION RESULTS

To understand the effect of OFDM on transmitted symbol, coding is done for SISO-OFDM and plotted the BER vs. SNR graph. For the simulation 51200 bits are transmitted during each iteration, size of FFT taken is 512, cyclic prefix length is 10 and channel length is 6. The simulation result is shown in figure 4.

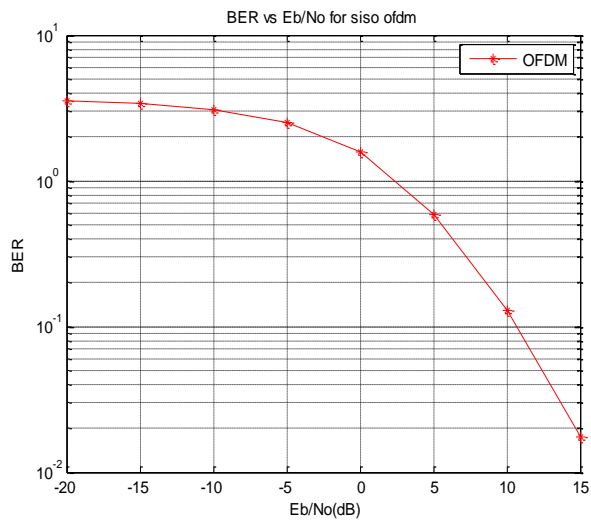

Figure 5. BER performance for SISO OFDM

SFBC for two transmit and receive antennas using LS and MMSE linear detection is simulated, considering 51200 bits transmitted for each SNR values and taking channel effect and AWGN, the result obtained is shown in figure 5.

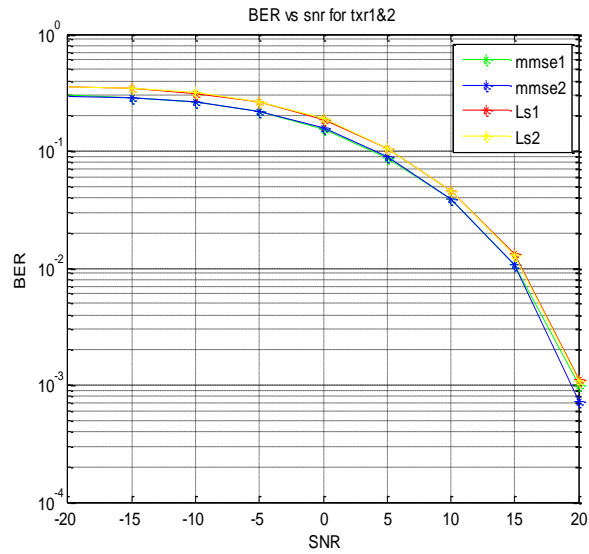

Figure 6. BER performance for 2x2 MIMO using SFBC MMSE and LS comparison

Comparing SNR value and corresponding BER value of figure 4 and figure 5 conclude that MIMO shows better performance. In figure 5, BER vs. SNR for LS and MMSE is compared; better performance is shown by MMSE. Further the effect of channel imperfections on BER is evaluated and result is that as the channel imperfection increases BER also increases. Channel is assumed to be constant during entire transmission but noise variance is kept increasing hence the effect of channel imperfection achieved and simulated the result, shown in figure 6.

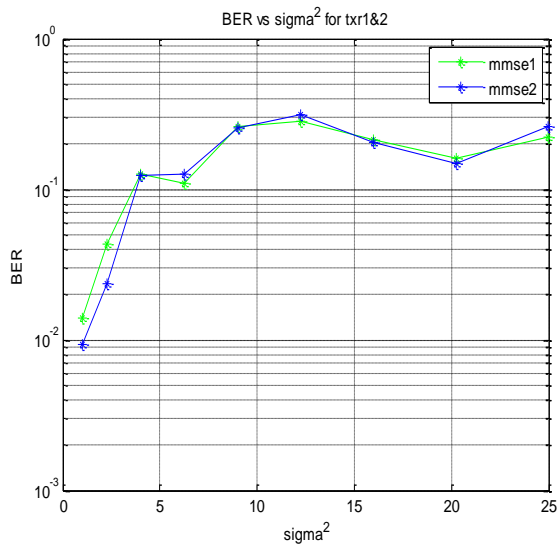

Figure 7. Channel imperfections in MIMO SFBC MMSE linear detection

## 7. FUTURE WORKS

For effective utilization of bandwidth, several users have to share same bandwidth. This leads to the evolution of multi user MIMO (MU-MIMO). When different users share same spectrum co-channel interference occurs and the effects needs to be mitigated. Several approaches like interference avoidance and treating interference as noise have several limitations, so to overcome that and to manage interference, interference alignment techniques will be introduced, adaptive beamformer design for interference alignment and cancellation its performance estimation and comparison of various beamformers will be evaluated under perfect and imperfect channel state information (CSI) in MATLAB tool [6].

## 8. CONCLUSION

As wireless communication is concerned, effects of channel and noise on transmitted signal are to be considered during the detection procedure, knowledge of channel variations or methods to overcome its effects is essential for a designer to predict and achieve higher efficiency. In this paper channel effects are considered and MMSE, LS linear detection methods are adopted to recover original data. Simulation results are plotted and results obtained, proves MMSE works best.

**Authors**

Gopika k has graduated from Ilahia College of Engineering and Technology of Mahatma Gandhi University in Electronics & Communication Engineering in 2014. She is currently pursuing her M.Tech Degree in Wireless Technology from Toc H Institute of Science & Technology, Arakunnam. Her research interest includes Signal Processing and Wireless communication. 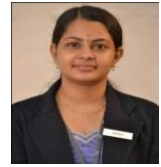

M. Mathurakani has graduated from AlagappaChettiar College of Engineering and Technology of Madurai University and completed his masters from PSG college of Technology of Madras University. He has worked as a Scientist in Defence Research and development organization (DRDO) in the area of signal processing and embedded system design and implementation. He was honoured with the DRDO Scientist of the year award in 2003.Currently he is a professor in Toc H Institute of Science and Technology, Arakunnam. His area of research interest includes signal processing algorithms, embedded system modeling and synthesis, reusable software architectures and MIMO and OFDM based communication systems. 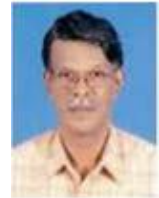